# Highly efficient water purification by $WO_3$-based homo/heterojunction photocatalyst under visible light

Kunihiko Kato [a] and Takashi Shirai [a*]

[a] Advanced Ceramics Research Center, Nagoya Institute of Technology, Gokiso, Showa-ku, Nagoya, Aichi 466-8555 Japan

## Abstract

Bandgap engineering has attracted increasing interest in the field of semiconductive materials with specific electrical and optical characteristics for various applications, such as solar energy harvesting, energy-related applications, and photocatalysts. Here, we propose a novel and facile method to obtain a combination of type-II structure and Ohmic contact by the fabrication of $WO_3$/Ti-$WO_x$/$TiH_yO_z$ homo/heterojunction on a $WO_3$ photocatalyst using ball milling. The morphology and structural characterizations of the material are obtained using UV–vis, SEM/EDS, XRD, Raman spectroscopy, and XPS. It is revealed that titanium hydride ($TiH_2$) facilitates simultaneous $WO_3$ reduction and cation doping, which generates intervalent W ions ($W^{5+}$, $W^{4+}$) with oxygen deficiency, leading to the formation of new band structures at the $WO_3$/Ti-$WO_x$/$TiH_yO_z$ interface. As a result, the homo/heterojunction significantly extends light absorption above 400 nm and facilitates the spatial separation of photoexcited carriers, which significantly enhances the photocatalytic performance to degrade azo-dye water pollutants under visible light irradiation.

[*] Corresponding author. E-mail: shirai@nitech.ac.jp

## 1. Introduction

Tungsten oxides (WO$_x$, x ≤ 3) have attracted considerable research interest because of their easy preparation, diversely tunable stoichiometry, crystal structure/particle morphology, bandgap (2.6 eV), strong photocatalytic oxidation, nontoxicity, and abundance [1]. Moreover, WOx-based materials have been applied to various fields, such as solar energy harvesting [2], gas sensors [3], energy-related applications (e.g., supercapacitors, batteries, and fuel cells) [4–6], and photocatalysis [7–9]. Compared to other semiconductor photocatalysts (e.g., TiO$_2$ and α-Fe$_2$O$_3$), WO$_x$ possesses higher carrier mobility and longer hole diffusion length [1,10,11]. However, bare WO$_x$ exhibits a relatively low efficiency owing to a poor electron-scavenging pathway, which increases the recombination rate of photoexcited carriers [1]. Thus, many researchers have used bandgap engineering strategies for efficient consumption of electrons by controlling the atomic configurations of crystals [9,12], controlling the stoichiometry [13,14], elemental doping [15], compositing with noble metal nanoparticles or carbon materials as electron mediators [16,17], and combining with other semiconductors [18] to form heterojunctions.

In particular, controlling the stoichiometry and introducing oxygen vacancies is a practical approach to extend light absorption and facilitate carrier transfer/separation, which significantly improves photocatalytic performance [1]. In the last decade, various methods were used to prepare nonstoichiometric tungsten trioxides (WO$_{3-x}$), such as the hydrothermal/solvothermal reaction of inorganic precursors or W [13,16], annealing WO$_3$ in a reducing atmosphere (H$_2$) or vacuum [7,19], and microwave plasma ablation of W in aqueous solution [20]. However, the reported approaches require unique experimental setups for thermal treatment in a vacuum or a hydrogen-rich atmosphere, high-pressure control, and introduction of specific reaction fields.

In this study, we propose a novel and facile bandgap engineering method to fabricate homo/heterojunctions on WO$_3$ at room temperature using ball milling to design a highly efficient visible-light photocatalyst (**Figure 1**). Here, titanium hydride (TiH$_2$) is significant to facilitate the following simultaneous reactions: (1) element doping and oxygen abstraction from WO$_3$ as a reducing agent and formation of specific electronic contacts at the interface of the hybrid material ($WO_3$/Ti-$WO_x$/$TiH_yO_z$); (2) easy tunability of the chemical composition in Ti-$WO_x$ in the range of $1.39 < x < 2.75$, bandgap narrowing, and valence band (VB) edge tailing by controlling the amount of $TiH_2$. As a result, the $WO_3$-based photocatalyst with $WO_3$/Ti-$WO_x$/$TiH_yO_z$ homo/heterojunction demonstrates a wide range of light absorption above 400 nm and superior spatial carrier separation, which results in the excellent photocatalytic performance (more than three orders of magnitude higher than that of pristine $WO_3$) for the degradation of azo-dye water pollutant under visible light irradiation. Furthermore, the synergistic effect of the homo/heterojunction structure on the enhancement of photocatalytic activity is discussed in this study.

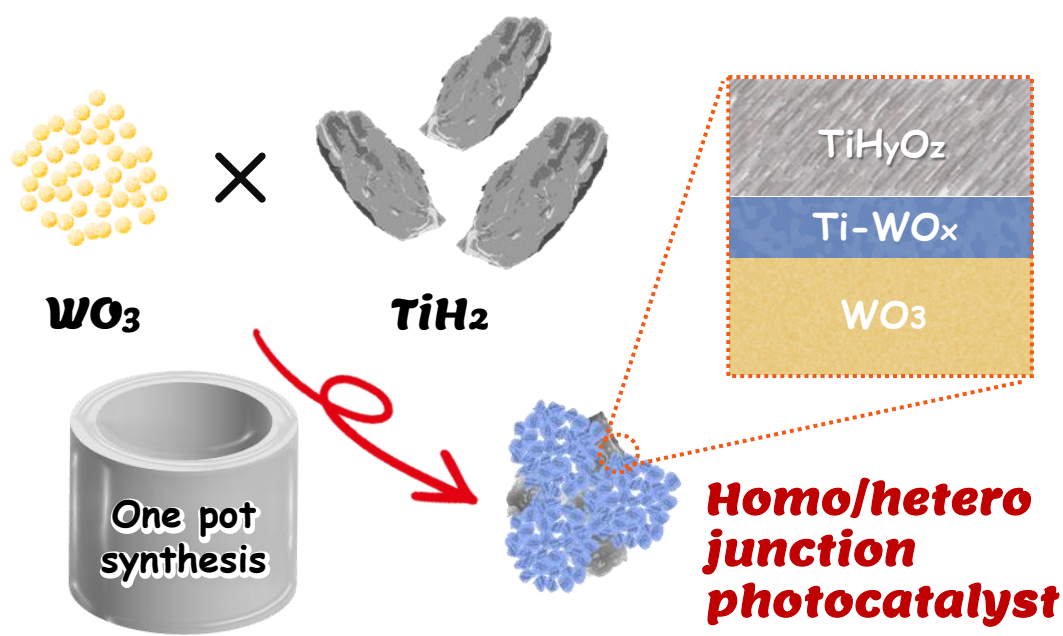

**Figure 1.** A facile and scalable buildup of homo/heterojunction $WO_3$ photocatalyst by the one-pot reaction.

## 2. Experimental

Commercially available pure $WO_3$ powder (>99.5%, Nacalai Tesque Co., Ltd) and $TiH_2$ (>99.9%, Kojundo Chemical Laboratory Co., Ltd.) powder were used as raw materials. High-energy ball milling was performed in a commercial planetary ball mill (Pulverisette-5, Fritsch GmbH) using zirconia balls of 5 mm diameter and zirconia pot of 80 ml. The ball milling was performed at a rotation speed of 300 rpm for 1 h. The mixture powder (volume ratio of $WO_3$ and $TiH_2$; 100-x:x, where x = 0, 1, 2, 4, and 8) was placed into the pot in a $N_2$-filled glovebox before the milling. The synthesized samples were labeled as W-Ti1, W-Ti2, W-Ti4, and W-Ti8, respectively.

The optical properties in the UV to visible range and UV to near-infrared range were analyzed by UV–vis spectrophotometry (V-7100, JASCO. The particle morphology was analyzed using field-emission scanning electron microscopy (JSM-7000F, JEOL). X-ray diffraction (XRD) patterns were obtained using an X-ray diffractometer (Ultima-IV, Rigaku). Raman spectroscopy was performed using a Raman spectrometer (NRS-3100, JASCO, or Via Raman microscope, Renishaw plc). The chemical state of the surface of the initial sample was examined using X-ray photoelectron spectroscopy (XPS; M-probe, Surface Science Instrument) with Al Kα source (hν = 1486.6 eV). The electron spin resonance (ESR) spectra were obtained using a JEOL JES-FA200 spectrometer. The Fourier transform infrared (FT-IR) spectra were obtained using an FT-IR spectrometer (FT-IR 6600, JASCO).

The photocatalytic performance was studied by degrading methyl orange (MO) as an azo-dye water pollutant under visible light. First, 1 mg of either raw $WO_3$ or the as-prepared powder was transferred into a vessel containing 2 mL MO solution at 15 ppm. A 300 W Xe lamp (MAX-350, ASAHI SPECTRA, Japan) with a band-pass filter (385–740 nm) was used as the visible-light source. Prior to irradiation, the solution containing the $WO_3$ photocatalysts was placed in the dark for approximately 3 h to ensure the equilibrium adsorption of the dye molecules. The MO degradation was monitored by examining the changes in the UV–vis absorption spectra at regular intervals after light irradiation. The photoluminescence (PL) spectra were obtained using a commercial spectrofluorometer (FP-8500, Jasco, Japan) at an excitation wavelength of 350 nm.

## 3. Results and Discussion

### 3.1. Optical property

WO$_3$-based hybrid materials exhibit a wide range of visible-light absorbing properties above 400 nm, as illustrated in **Figure 2a**. The absorbance increases with the increasing volume fraction of TiH$_2$ powder, and the absorption band edge gradually shifts toward longer wavelengths (redshift). The wide-range light absorption and edge shift are commonly observed in WO$_x$ [1,7,21] owing to the weakening of the W–O interaction by the Ti doping [23]. The optical bandgaps are determined from the slope in the Tauc plot, as shown in **Figure 2b** (the inset summarizes the values as a function of the volume fraction of TiH$_2$). The bandgaps of the composites narrow from 2.56 to 2.23 eV, which correspond to the reported optical bandgap of WO$_{3-x}$ [1,7,21]. Such bandgap narrowing is generally observed in WO$_x$ [23], where new energy levels are localized below the conduction band minimum (CBM) of bulk WO$_3$ (2.6 eV) owing to the formation of intervalent tungsten compounds (i.e., W$^{5+}$ and W$^{4+}$) with oxygen vacancies [22]. The trapped electrons in the W$^{5+}$/W$^{4+}$ states polarize the WO$_3$ lattice to induce polarons, which are associated with localized surface plasmon resonance (LSPR) [10]. Thus, the strong absorption in the visible and NIR regions can be attributed to the LSPR. Moreover, the high-energy ball milling of the yellow-green WO$_3$ powder with TiH$_2$ results in a dark-blue mixture (**Figure 2c**).

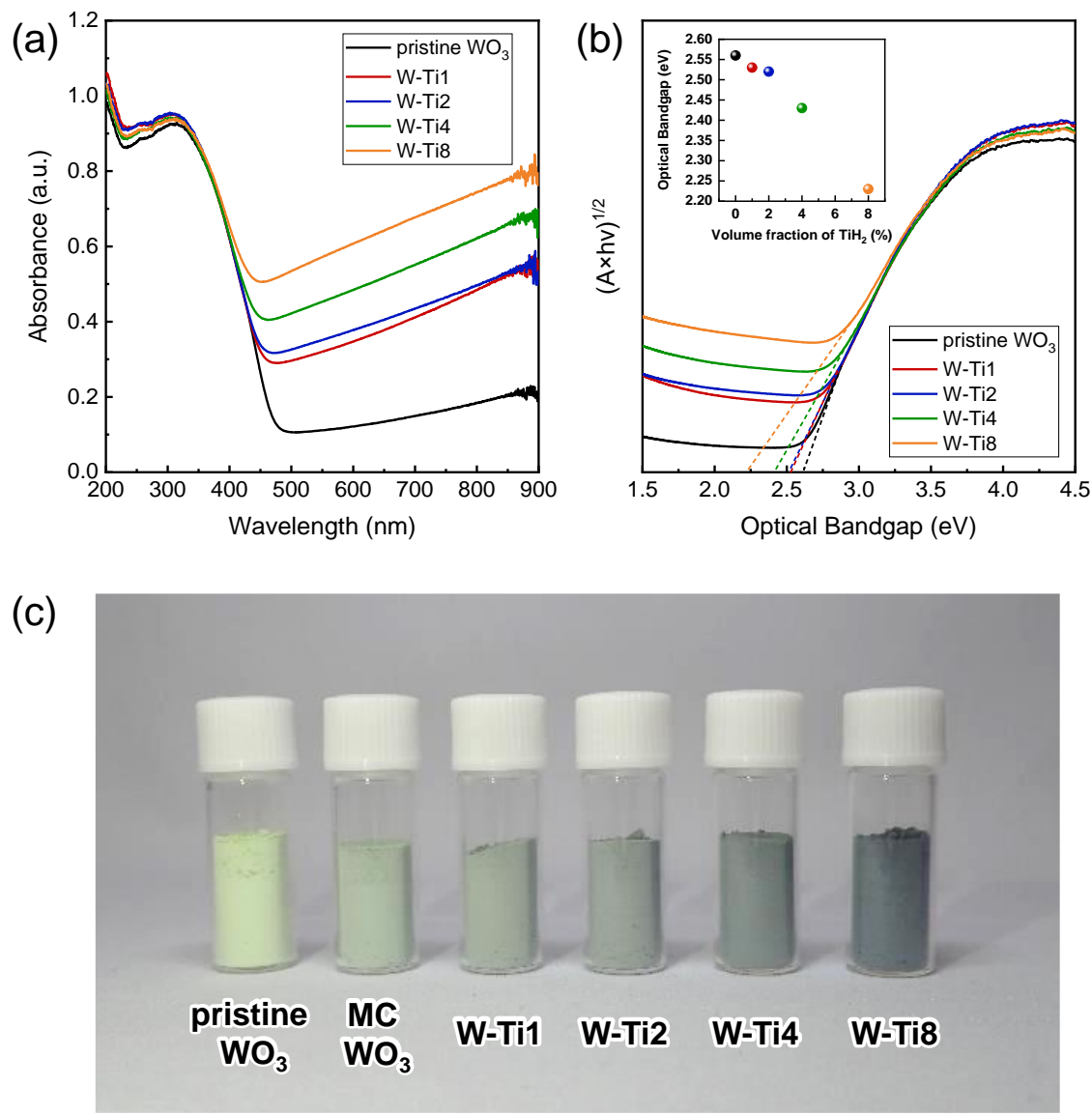

**Figure 2.** Optical property; (a) UV-vis spectra, (b) Tauc-plot, and (c) photograph of samples.

3.2. Particle morphology

Pristine WO$_3$ consisted of 40–50 nm coarse micron-sized particles. However, the mechanical treatment changes the morphology owing to the formation of neck and adhesion, as shown in **Figure S1**. In addition, the TiH$_2$ particles with angular shapes were not observed after the milling because they were embedded in the WO$_3$ agglomerates. **Figure S2** shows the nitrogen adsorption/desorption isotherms, which are all categorized as type II [24]. The Brunauer–Emmett–Teller (BET) specific surface area increased from 6.11 to 16.46 m$^2$g$^{-1}$. The synthesized composites exhibited a uniform distribution of Ti, as characterized by SEM-EDS (**Figure S3**). Aggregates of size 2–5 μm (the brighter points) indicated the presence of titanium compounds obtained from TiH$_2$. In this case, the mechanical treatment resulted in the reduced crystallinity of the monoclinic WO$_3$ phase (COD card No. 2106382). However, the phase transformation of WO$_3$ and the generation of new crystalline titanium compounds were not observed in the XRD patterns (**Figure 3**).

The structural disorder was confirmed by the Raman spectra (**Figure 4**), according to which the amount of TiH$_2$ strongly depended on the broadening of two primary peaks at 710 and 810 cm$^{-1}$ (attributed to phonons in crystalline WO$_3$ [25–27]). The peak broadening could be ascribed to the oxygen deficiency in the WO$_3$ lattice [7]. To examine the change in the reduction of synthesized Ti-WO$_x$, the Raman maps were obtained at 60 × 60 μm$^2$ (all results are displayed in **Figure S4**). **Figure S5a** shows the histogram as a logarithmic function of the peak intensity at 810 cm$^{-1}$, which can be expressed by a normal distribution with a mean of μ. **Figure S5b** shows a good relationship between the log-scale μ and the volume fraction of TiH$_2$ in the mixture, suggesting that the reduction occurred with the Ti doping/oxygen abstraction at the WO$_3$/TiH$_2$ interface during high-energy ball milling.

The analysis of the chemical structure of the nanosurface by XPS revealed the reduction state of the Ti-WO$_x$ nanoparticles (**Figure 5**). In W4f XPS spectra, the shoulder peaks appeared at approximately 37.2/35.3 and 36.4/34.2 eV, which corresponded to W$^{5+}$ and W$^{4+}$, respectively [7]. The area fraction attributed to the intervalent W ions (W$^{5+}$ and W$^{4+}$) increased with the increase in the volume fraction of TiH$_2$. Moreover, OH groups and physisorbed H$_2$O molecule-derived shoulder signals [7,14] were detected, as shown in the O1s XPS spectra. In general, oxygen vacancies facilitate the dissociative chemisorption of water molecules in nonstoichiometric transition metal oxides [28]. In particular, the W–O–Ti chemical bond was observed in the W–Ti8 sample. In this case, every signal from Ti was detected by neither a survey nor a narrow scan (**Figure S6**), suggesting the presence of atomically dispersed Ti ions in the synthesized WO$_{3-x}$ nanoparticles. The chemical composition of Ti–WO$_x$ was tunable in the range 1.39 < x < 2.75. Moreover, the edge of the VB shifted toward lower binding energy from 1.56 to 0.08 with the increase in TiH$_2$. The elemental doping with the oxygen deficiency could tune the VB of WO$_3$ (O2p orbital) [15,29]. Recently, some research groups reported that oxygen deficiency led to the formation of new levels that partially overlapped with the VB [30,31]. This contributed to the VB tailing and bandgap narrowing, which extended the photo-response.

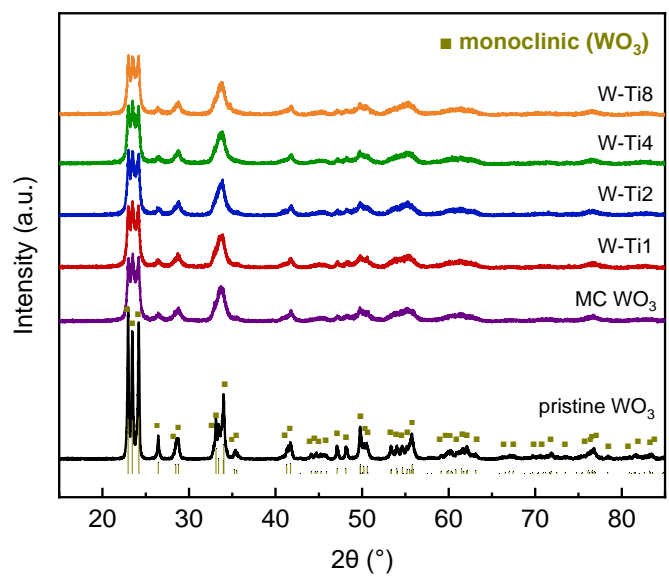

**Figure 3.** XRD patterns.

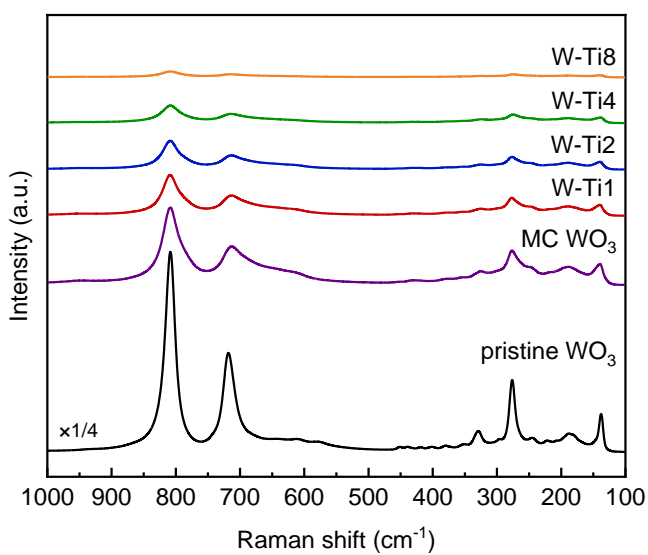

**Figure 4** Raman spectra.

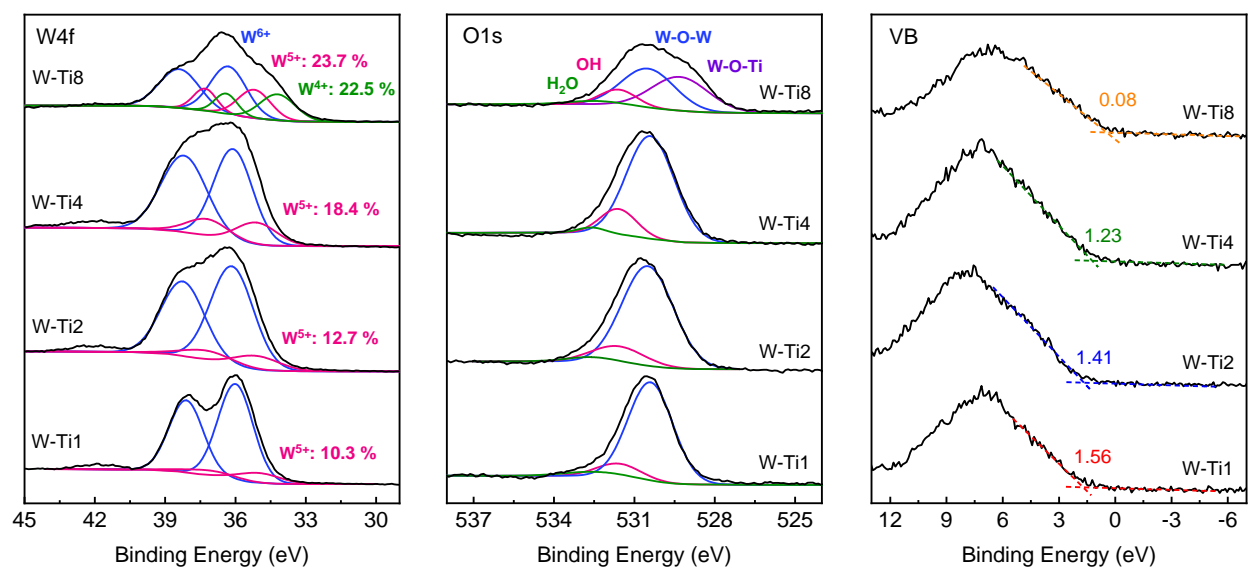

**Figure 5.** XPS spectra (W4f, O1s, and VB).

3.3. Visible-light photocatalytic activity

The visible-light photocatalytic activity of the $WO_3$-based hybrid nanoparticles is significantly enhanced for the photodegradation of the azo-dye water pollutant, methyl orange (MO; 15 ppm), as shown in **Figure 6a.** UV–vis spectra of the MO solution before the adsorption equilibrium and during the photodegradation are shown in **Figure S7**. The photocatalytic degradation kinetics follow the first-order law, expressed as $\ln(C_0/C) = kt$ [32], where t is the irradiation time and $k$ is the degradation rate constant. The values of $k$ determined from the slope of the linear curves in **Figure 6b**, ranged from < $1.00 \times 10^{-5}$ (pristine $WO_3$) to $2.54 \times 10^{-2}$ min$^{-1}$ (W-Ti8). Thus, the one-pot mechanical reaction enhanced the photocatalytic degradation rate by three orders of magnitude. **Table S1** compares the results of this study with the performance of other $WO_3$ and semiconductor photocatalysts, indicating that the photocatalytic performance of the synthesized photocatalyst is superior even with a lower surface area and earth-abundant elements. However, the excess loading of $TiH_2$ during the ball milling hinders further activity enhancement due to the severe deformation of the $WO_3$ lattice by forming a high concentration of oxygen vacancies. As shown in **Figure 6c**, the original MO displays strong absorption at 465 and 270 nm, which corresponds to the azo bond (-N=N-) and the benzene ring, respectively [33,34]. The main absorption peak (465 nm) shows a redshift before the adsorption equilibrium (photograph is shown in **Figure 6d**). Azo dye is commonly used as a pH indicator and MO changes the color from yellow to red with increasing acidity. The redshift indicates that the W–OH (in this case, OH groups formed on the $WO_x$ surface, as shown in the XPS results) have Brønsted acidic features [35], which function as the active sites for the photocatalytic reaction. Some research groups have reported that the Brønsted acid sites in semiconductor photocatalysts are significant for activity enhancement (e.g., esterification [36] and degradation of MO [37]). The intensity of the central peak decreased quickly after the visible-light irradiation owing to the cleavage of the conjugated double bond with decolorization by photocatalytic degradation. Moreover, several degraded products were detected, which can be ascribed to amines (<240 nm) [38], sulfanilic acid, and N, N-dimethyl-p-phenylenediamine (240–260 nm) [33,34], by the photodegradation. The recombination rates of the photoexcited carriers were obtained by the PL measurements (Figure S8), where the PL intensity decreased after compositing. The results indicate that the homo/heterojunction formation was significant for enhancing the carrier separation.

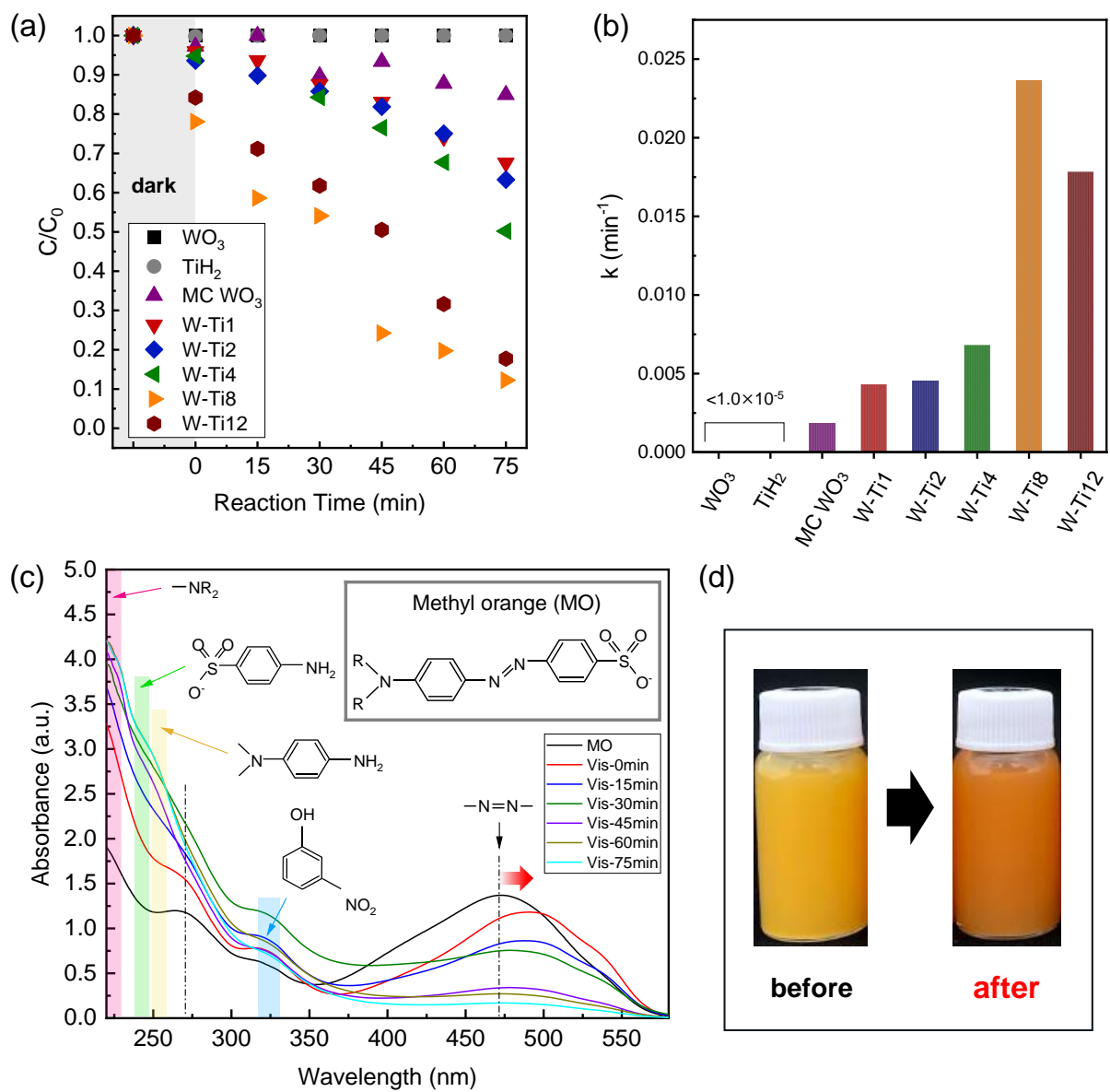

**Figure 6.** (a) Photocatalytic performance for the degradation of MO (15 ppm) under visible light irradiation, (b) photocatalytic reaction rate constant, (c) UV-vis spectra of MO solution during photodegradation (W-Ti8), and (d) photographs of MO solution before and after adsorption equilibrium.

3.4. Influence of homo/heterojunction on photocatalytic activity

To investigate the influence of the $WO_3$/Ti-$WO_x$/$TiH_yO_z$ homo/heterojunction on the photocatalytic performance, an additional chemical treatment was performed using NaOH to eliminate a part of the $WO_3$ phase. The obtained material mainly consisted of $TiH_2$ crystals, and the XRD peaks shifted to lower 2-theta values, which could be attributed to the lattice shrinkage caused by the oxidation of $TiH_2$ [39] (**Figure 7a**). However, in the Raman spectroscopy (**Figure S9**), the signals originating from $WO_3$ were detected even after alkaline dissolution. In addition, the XPS measurements revealed the presence of intervalent Ti ions and W–O–Ti in the Ti2p and O1s XPS spectra, respectively (**Figure 7b**). Thus, the results suggested that the strong interaction with $TiH_yO_z$ at the interface chemically stabilized Ti–$WO_{3-x}$. **Figure 7c** displays the EDS mapping of the chemically treated particles, where W is uniformly distributed in the $TiH_yO_z$ particles. Furthermore, the remaining Ti–$WO_x$ nanoparticles exhibited a superior light-absorbing capacity in the visible region and highly efficient spatial carrier separation compared to the chemically untreated particles (W–Ti8) (**Figure 7d,e**).

**Figure 8** shows the schematic illustration of a possible mechanism for enhancing the photocatalytic activity by the $WO_3$/Ti–$WO_x$/$TiH_yO_z$ homo/heterojunction on the $WO_3$ nanoparticles. Generally, $WO_3$ with oxygen deficiency behaves as an n-type semiconductor with the Fermi level positioned 0.2–0.3 eV below the CBM [40,41]. Moreover, the electron affinity of $WO_3$ is 5.2–5.3 eV [41]. However, the work function of Ti is 4.3–4.4 eV [42], and this value decreases by ~250 meV by the hydrogen uptake ($TiH_x$, $x \leq 2$) [43]. In addition, $TiH_2$ exhibits a metallic feature with an electrical conductivity comparable to that of titanium [32]. Based on the upward tailing of the valence band maximum (VBM) up to 1.5 eV and the optical bandgap narrowing (2.23 eV for W-Ti8) in the formed Ti-$WO_x$, the CBM is located at approximately 5.0 eV relative to the vacuum level, which has a higher energy potential than $TiH_2$ ($TiH_yO_z$). A similar value of the Fermi level in $WO_3$ is assumed in Ti-$WO_x$, as shown in the report for transition metal-doped $WO_3$ [44]. Thus, in the heterojunction, the photogenerated electrons on Ti-$WO_x$ can be easily transferred to the metallic $TiH_yO_z$ side by Ohmic contact, which significantly inhibits carrier recombination. Such Ohmic contacts were observed at the interface between the transition metal and n-type semiconductors (e.g., W/WO$_3$ [45], Fe/Fe$_2$O$_3$ [46]). Moreover, the prepared nanocomposites consist of type-II [47] homojunction. As charge carrier transfer in the type-II structures, the photo-excited electrons will accumulate on the CB of Ti-WO$_x$ with weak reduction potential. In contrast, the photo-excited holes will gather on the VB of WO$_3$ with weak oxidation potential, resulting in the spatial charge separation. Considering the possible mechanism of the MO degradation based on previous reports [48,49], the following reactions occur in the synthesized composite photocatalyst:

$$WO_3 \text{ or } Ti\text{-}WO_x \rightarrow h_{VB}^+ + e_{CB}^- \quad (1)$$

$$O_2 + e_{CB}^- \rightarrow O_2^{\bullet -} \quad (2)$$

$$O_2^{\bullet -} + 2 H^+ + e_{CB}^- \rightarrow H_2O_2 \quad (3)$$

$$4H_2O + 2 h_{VB}^+ \rightarrow H_2O_2 + 2H^+ \quad (4)$$

$$H_2O_2 \rightarrow 2 \bullet OH \quad (5)$$

$$\bullet OH/ O_2^{\bullet -} + MO \rightarrow \text{Degradation products} \quad (6)$$

In general, the excited electrons in the CB of bulk WO$_3$ have less potential to reduce O$_2$ to superoxide anion radicals (O$_2^{\bullet -}$). However, in this case, many photoexcited holes and electrons could be retained on different counterparts with superior light-absorbing capacity, which enhances the photocatalytic efficiency by a combination of type-II structure and Ohmic contact in the WO$_3$-based homo/heterojunction photocatalyst. As a further experiment, the predominant active species generated in the photocatalytic reaction were detected by scavengers. Adding EDTA-2Na and t-BuOH restrain the photocatalytic reaction significantly (**Figure S10**), indicating that h$^+$ and •OH are the predominant active species for MO degradation. Therefore, the WO$_3$/Ti–WO$_x$ homojunction plays an essential role in the significant activity enhancement.

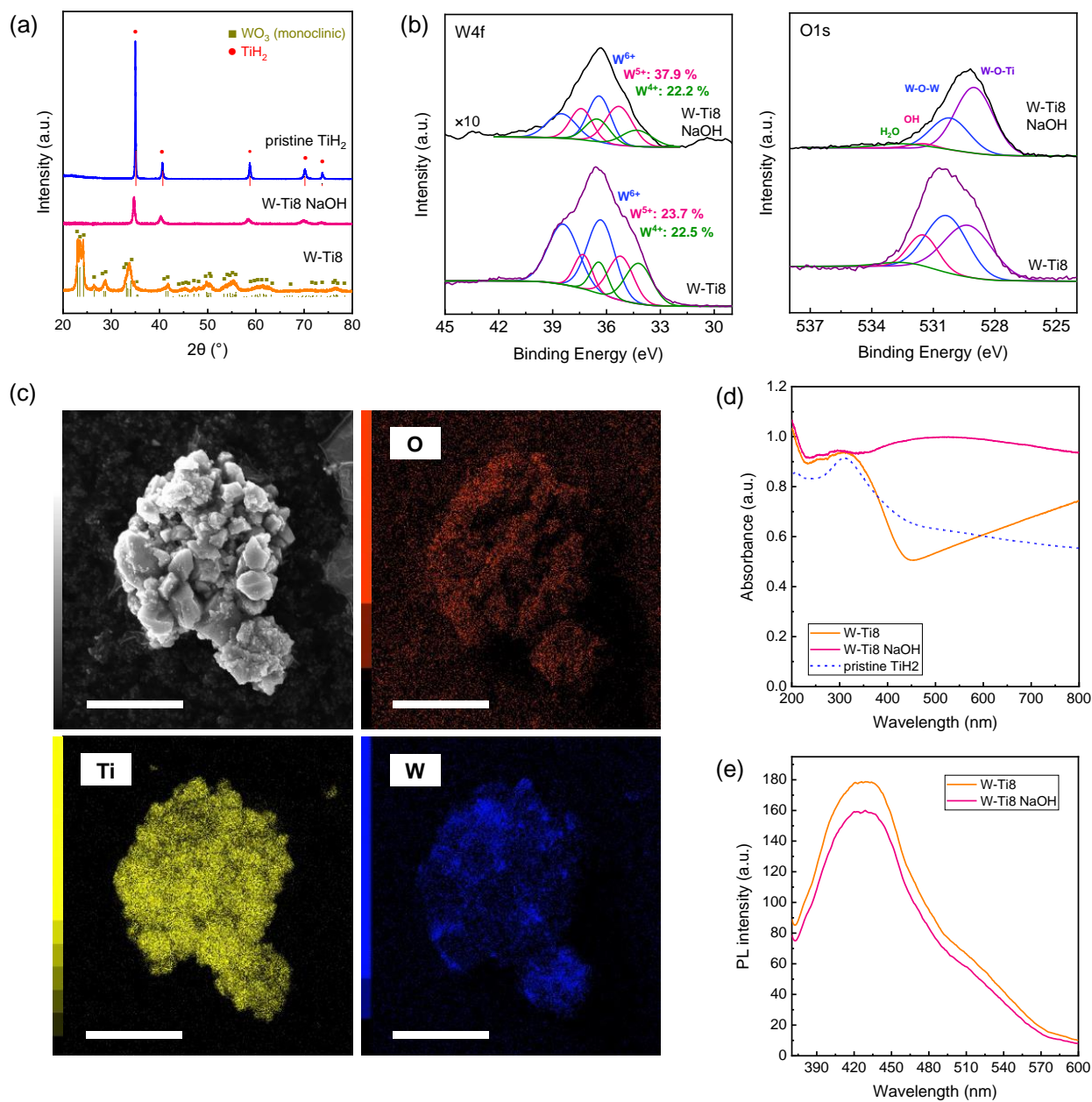

**Figure 7.** Characterization of the obtained material after chemical etching; (a) XRD patterns, (b) XPS spectra (Ti2p and O1s), (c) EDS mapping (scale bar = 25μm), (d) UV-vis spectra and (e) PL spectra.

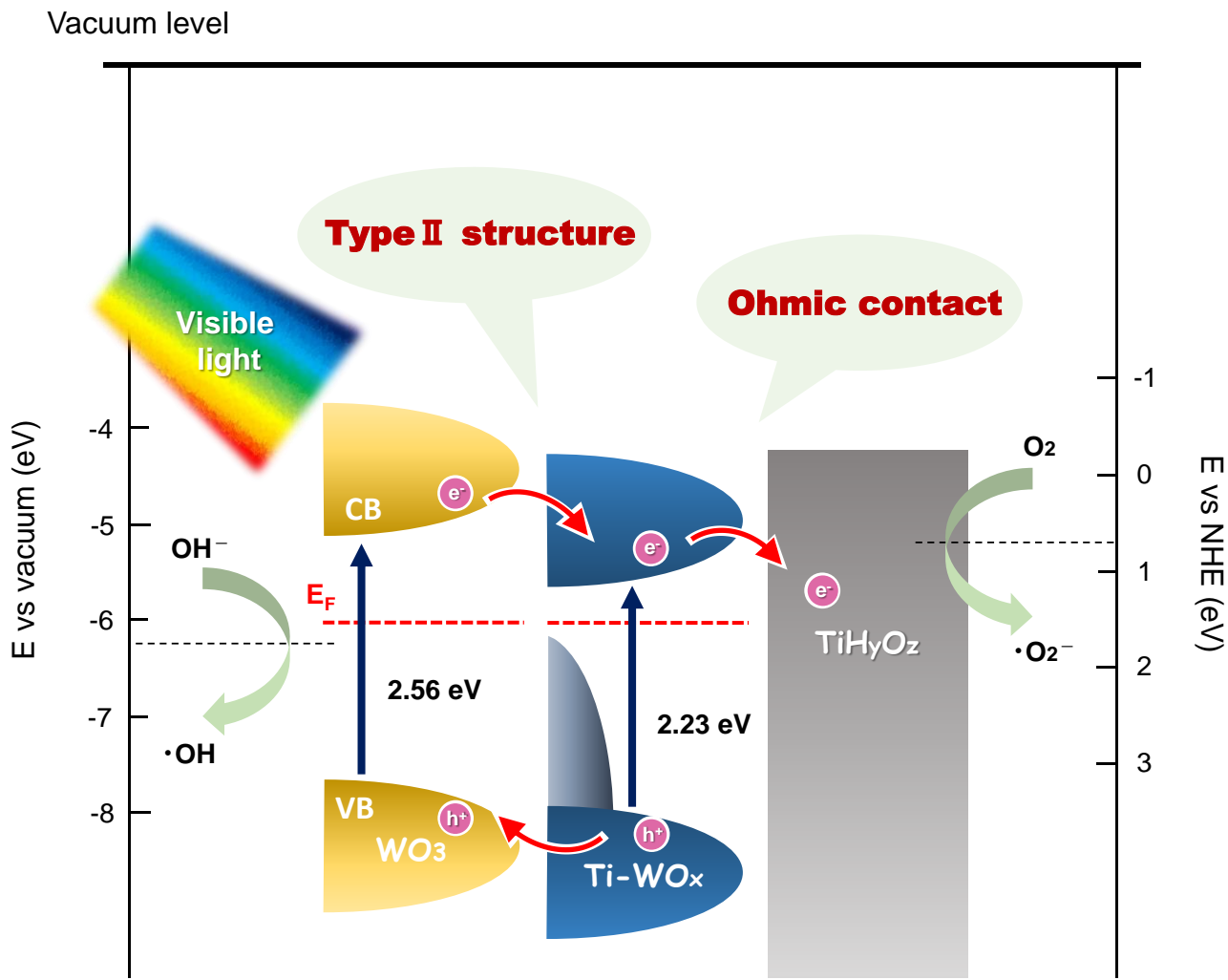

**Figure 8.** Schematic illustration for the mechanism of photocatalytic activity enhancement by a hybrid strategy as a combination of type-II structure and Ohmic contact in homo/heterojunction $WO_3$ photocatalyst.

## 4. Conclusions

In summary, we proposed a novel and facile bandgap engineering method with a combination of type-II structure and Ohmic contact by the fabrication of the $WO_3$/Ti-$WO_x$/Ti$H_y O_z$ homo/heterojunction on the $WO_3$ nanoparticles. Ti$H_2$ facilitated simultaneous $WO_3$ reduction and cation doping by high-energy ball milling, which offered a wide range and strong light absorption above 400 nm. Furthermore, the chemical composition of Ti-$WO_x$ was tunable in the range $1.39 < x < 2.75$ by controlling the amount of Ti$H_2$. In this case, the Ti element doping and oxygen abstraction by Ti$H_2$ generated the intervalent W ions ($W^{5+}$, $W^{4+}$) with oxygen deficiency, which resulted in the upward shift of the VB edge up to 1.5 eV and bandgap narrowing. Furthermore, the formed Ti-$WO_x$ phase exhibited high chemical stability owing to the formation of strong interactions at the interface with Ti$H_y O_z$ as an Ohmic contact (heterojunction). Furthermore, many photoexcited holes and electrons could be retained on different counterparts because the hole on Ti-$WO_x$ recombined with the excited electron from the $WO_3$ homojunction as a type-II structure (homojunction). The homo/heterojunction facilitated the spatial separation of the photoexcited carriers, which significantly enhanced (more than three orders of magnitude higher than that of pristine $WO_3$) the photocatalytic performance for the degradation of the azo-dye water pollutant (MO) under the visible-light irradiation.